# Gas Sensing Properties of Single Conducting Polymer Nanowires and the Effect of Temperature

Yaping Dan[1], Yanyan Cao[2], Tom E. Mallouk[2], Stephane Evoy[3], A. T. Charlie Johnson[1,4]

[1]Department of Electrical Engineering, University of Pennsylvania, Philadelphia, PA 19104, USA,

[2]Department of Chemistry, Pennsylvania State University, College Park, PA 16802, USA

[3]Electrical and Computer Engineering and National Institute of Nanotechnology, University of Alberta, Edmonton, AB T6G 2V4 Canada,

[4]Department of Physics and Astronomy, University of Pennsylvania, Philadelphia, PA 19104, USA

**Abstract**

We measured the electronic properties and gas sensing responses of template-grown poly(3,4-ethylenedioxythiophene)/poly(styrenesulfonate) (PEDOT/PSS)-based nanowires. The nanowires have a "striped" structure (gold-PEDOT/PSS-gold), typically 8μm long (1 μm – 6 μm – 1 μm for each section, respectively) and 220 nm in diameter. Single-nanowire devices were contacted by pre-fabricated gold electrodes using dielectrophoretic assembly. A polymer conductivity of 11.5 ± 0.7 S/cm and a contact resistance of 27.6 ± 4 kΩ were inferred from measurements of nanowires of varying length and diameter. The nanowire sensors detect a variety of odors, with rapid response and recovery (seconds). The response (ΔR/R) varies as a power law with analyte concentration.

The development of a low footprint versatile "electronic nose" (e-nose) system will open a wide range of applications such as clinical assaying, emission control, explosive detection, and workplace hazard monitoring.[1-3] An e-nose system[4] consists of an array of odor sensors and a computational system to convert the pattern of sensor responses elicited by exposure to a given volatile analyte into a computed response reporting recognition and categorization of the analyte[5,6]. Responses from the sensor array produce a combinatorial code for each volatile analyte, as in biological olfaction.[7,8] The sensor array ideally exhibits the range of selectivity and sensitivity to volatile analytes displayed by biological olfactory receptors,[9-15] although this has not yet been achieved. One approach to such an e-nose system would entail integrating an extremely large sensor array with CMOS signal-processing circuitry. Template-grown metal nanowires were previously integrated with pre-fabricated CMOS circuitry using dielectrophoretic assembly.[16] The desire to fabricate a very dense array of discrete receptors for volatile analytes suggests the use of nanoscale devices, in particular nanowire sensors, which typically exhibit performance advantages due to their large surface-to-volume ratio and quasi-one-dimensional electronic transport. CP vapor sensors respond to a wide range of analytes, and the sign and magnitude of the response depends on the choice of polymer[17-20]. These considerations motivate the investigation of CP nanowire vapor sensors for use in an e-nose sensor array.

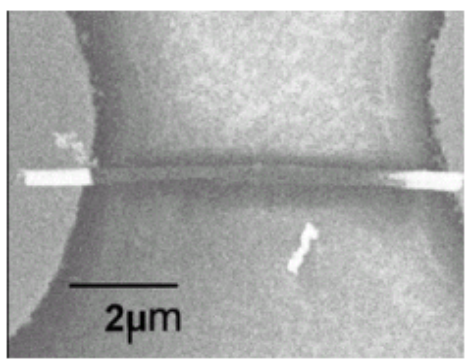

Figure 1. SEM image of a striped nanowire assembled onto a pair of gold electrodes.

Poly(3,4-ethylenedioxythiophene) /poly(styrenesulfonate) (PEDOT/PSS) is a particularly stable conducting polymer that has received sustained interest in recent years.[21,22] Here we report on the electronic properties of single PEDOT/PSS nanowires integrated into arrays using dielectrophoretic assembly. We have also measured their gas-sensing responses and how these characteristics change with temperature.

In order to establish effective contacts with gold electrodes, the nanowires were synthesized with a "striped" structure (gold-polymer-gold) using a nanoporous template and multiple electrodeposition steps[23]. The striped nanowires were then released from the template and dielectrophoretically assembled onto prefabricated gold electrodes to yield a nanowire array[23] (Figure 1). The contact resistance between gold and polymer portions of the nanowire was intrinsic and relatively small since they were electrochemically synthesized. The two gold ends of the nanowire yielded an excellent, reproducible contact with prefabricated gold electrodes.

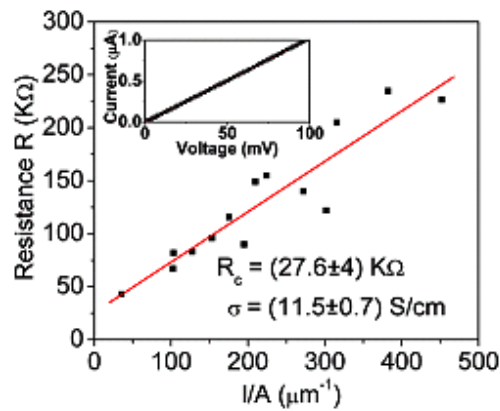

Figure 2. Resistance vs. length/(cross-section) l/A is plotted for 14 samples. Inset: Current-voltage characteristic of a single polymer nanowire.

The devices typically had a linear current-voltage (I-V) characteristic with resistance of order 100 k$\Omega$ (inset of Fig. 2). In order to extract the device contact resistance $R_c$ and the electrical conductivity of the polymer, we measured the resistance of 14 samples with varying diameter and length of the polymer region. Assuming the contact resistance $R_c$ and PEDOT/PSS electrical conductivity $\sigma$ are constant, the sample resistance $R$ should be given by $R = R_c + l/\sigma A$, where $l$ is the length of the polymer part of the wire, and $A$ its cross-sectional area, both measured using Scanning Electron Microscopy (SEM) and Atomic Force Microscopy (AFM). We find that the polymer portion of the nanowire is typically 6 ± 1 μm long and 220 ± 20 nm in diameter.

Figure 2 is a plot of R vs. $l/A$ for 14 samples, demonstrating that these quantities are linearly correlated as expected, with the contact resistance and PEDOT/PSS conductivity found to be $R_c = 27.6 \pm 4$ k$\Omega$, and $\sigma = 11.5 \pm 0.7$ S/cm, respectively. The contact resistance consists of two parts: the contact resistance between the polymer and gold caps and the contact between the gold caps and gold electrodes. We verified that the second contribution is negligible (less than 100 $\Omega$) by measuring the electrical resistance of pure gold nanorods assembled onto gold electrodes using the same technique.

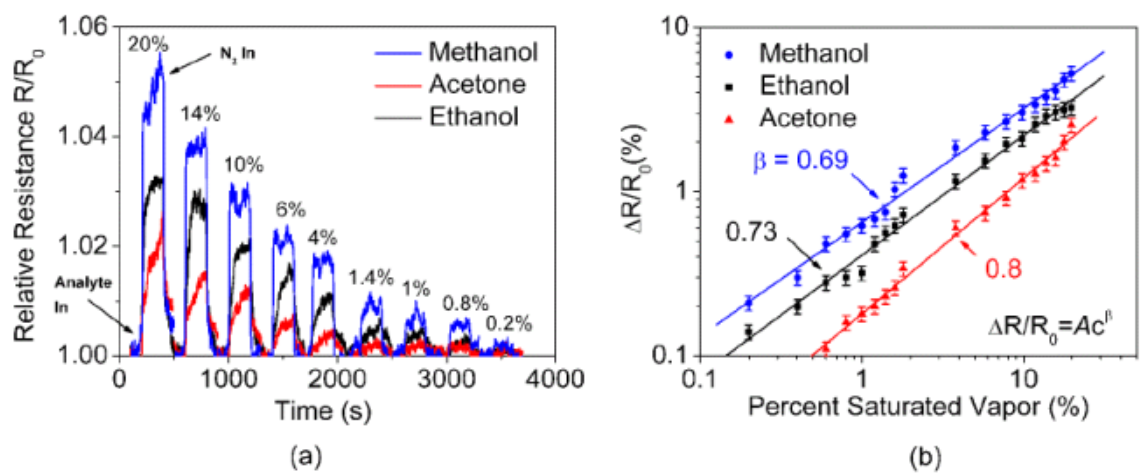

Figure 3. Nanowire sensor responses to methanol, ethanol and acetone. (a) The response in time to different analyte concentration, given as a fraction of the saturated vapor. (b) Fractional increase in device resistance as a function of concentration for various analytes.

CP nanowire sensors were exposed to methanol, ethanol and acetone vapor of various concentrations. Typical sensor response data are plotted in Fig. 3(a). The nanowire shows rapid (~30s), reversible responses to all three analytes, and rapid recovery to baseline when exposed to air. Sensor response ($\Delta R/R$) as a function of analyte concentration c (Fig. 3b) shows a power-law variation: $\Delta R / R = Ac^\beta$. The exponent β is found to increase with the molecular weight of the analyte: the values of β are 0.69, 0.73, 0.80 for methanol, ethanol and acetone, respectively. We defined a minimum detectable concentration for a single device based on the noise floor ($\Delta R / R \sim 0.1\%$), which was intrinsic to the nanowires. The detection limit was found to be approximately 0.06% (76ppm), 0.14% (110ppm), and 0.5% (1200ppm) of a saturated vapor for methanol, ethanol and acetone vapor, respectively. The electrical response of the nanowire sensor is approximately 10 times faster than that reported for PEDOT/PSS film sensors,[24, 25] with comparable sensitivity. Additional experiments are required to determine whether smaller diameter devices offer further improvements over the thin film counterparts.

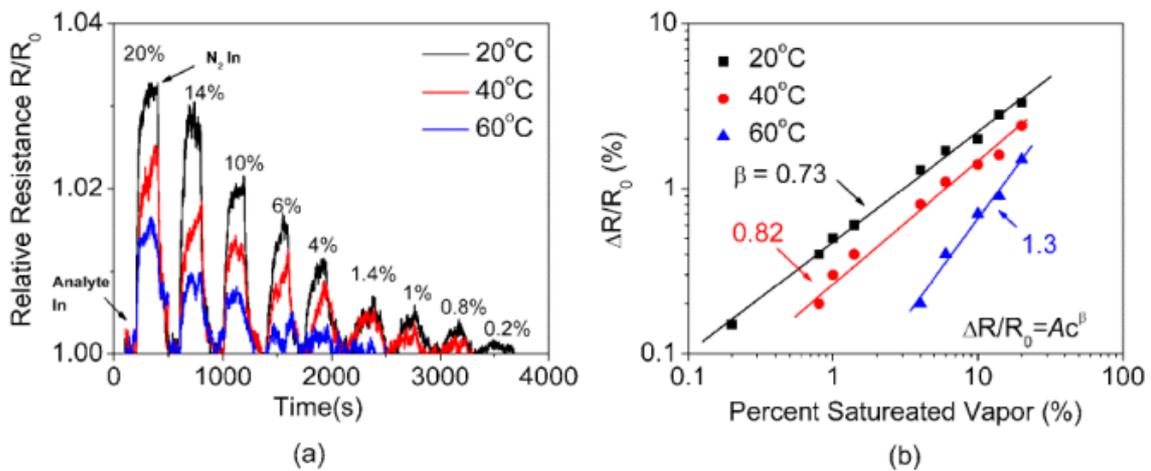

Figure 4. (a) Sensor response to ethanol vapor of various concentrations at elevated temperatures; (b) the fractional change in resistance vs concentration follows a power law.

The resistance of the nanowire devices decreased by about 8% over the temperature range 20 – 60 C. When exposed to ethanol vapor at elevated temperatures, the nanowire response to analytes again followed a power law, with the exponent increasing from 0.73 at 20 C to 1.3 at 60 C (Fig. 4). The response $\Delta R/R$ at a given concentration was a decreasing function of temperature, consistent with the expectation that analyte binding sites will be less occupied on average at higher temperatures.

It is remarkable that the sensor response shows a power law behavior as has also been reported for metal oxide vapor sensors.[26, 27] Such power laws have been explained theoretically in a model that incorporates both depletion of a semiconductor surface (grain boundary) and the chemistry of gas adsorption and reactions. The presence of a power law in this conducting polymer system suggests that this model or a related variation may be applicable to a far broader range of materials systems.

In summary, striped PEDOT/PSS nanowires (Au-polymer-Au) were electrochemically synthesized using the templating method and dielectrophoretically assembled with high yield onto pre-fabricated gold electrode pairs. Based on measurements of more than a dozen devices, the polymer electrical conductivity and gold-polymer contact resistance were found to be 11.5 ± 0.7 S/cm and 27.6 ± 4 kΩ, respectively. When exposed to vapors of organic analytes, the resistance of single nanowire devices followed a power-law variation $\Delta R / R = Ac^{\beta}$ as a function of vapor concentration. The power law exponent β was found to increase with molecular weight of the analyte and as a function of temperature.

**ACKNOWLEDGEMENTS**

This work was supported by the National Science Foundation under NIRT grant ECS-0303981, and the JSTO DTRA and the Army Research Office Grant # W911NF-06-1-0462.